\documentclass[prb,twocolumn,aps,floats,floatfix]{revtex4}
\usepackage{epsf,amsmath,amssymb}
\usepackage{graphicx}

\newcommand{\br}{{\boldsymbol r}}

\begin{document}
\setlength{\unitlength}{1cm}
\renewcommand{\arraystretch}{1.4}

\title{Island formation and dynamics of gold clusters on amorphous
  carbon films}

\author{Ralph Werner,$^1$ Matthias Wanner,$^2$ G\"unter Schneider,$^1$
  and Dagmar Gerthsen$^2$}  
\affiliation{$^1$Institut f\"ur Theorie der Kondensierten Materie,
  Universit\"at Karlsruhe, 76128 Karlsruhe
\\   
$^2$Laboratorium f\"ur Elektronenmikroskopie, Universit\"at
  Karlsruhe, 76128 Karlsruhe}


\date{\today}


\begin{abstract}
Samples of Au clusters deposited by laser ablation on an
amorphous-carbon substrate are investigated. After a few
months storage at room temperature the initially statistically
distributed clusters are found to be collected in agglomerates
consisting of larger clusters embedded in an Au film typically covering
areas of size 25$\times$70 nm$^2$. The Au film is determined to be
probably 4 to 8 monolayers but at most 7 nm thick. Evidence is found
that a number of clusters consisting of less then 50 atoms are pinned
at intrusions of the substrate. These results were derived using high
resolution transmission electron microscopy and off-axis holography
measurements to characterize the agglomerates as well as the
substrate. Monte-Carlo simulations were performed to model the film
formation process. To this end the substrate-Au interaction was
determined using density functional calculations (GGA) while the Au-Au
interaction was modeled with effective many body Gupta potentials. The
film formation can be understood as diffusion and fusion of clusters
of intermediate ($50< N < 300$ atoms) size. Larger clusters are more
stable at room temperature and remain adsorbed on the Au film.   
\end{abstract}

\maketitle


\section{Introduction}

The dynamic behavior of clusters deposited on a substrate has been
studied for several
decades.\cite{WG75,Will87,MRC96,PKW+01,EHD02,CPP03} A continuously
growing number of potential applications in
electronics\cite{FSCN04,ADKL02} and catalysis\cite{PLK+03,IAO+03}
demands defined arrays of nanoparticles. However, such arrays tend to
minimize their energy by minimizing the total surface via the growth
of larger particles at the expense of smaller ones.\cite{WG75} An
example for such a process is Ostwald ripening, where the larger
cohesive energy in larger particles leads to matter transport away
from small particles in the presence of a finite partial pressure of
the constituent in the environment surrounding the
clusters.\cite{Ostw00,Wagn61} Therefore, investigations concerning the 
stability\cite{Utla80,SPWB86} of deposited nanoparticles are of
considerable interest with respect to potential nano-technological
applications.  

Moreover, the thermodynamic\cite{ESZ+00,WDM+00} and
chemical\cite{HSH+03} properties of small metal clusters themselves
have enjoyed a large interest over the past years. Their properties
differ from those of the bulk material raising the fundamental
question about the statistical mechanics of finite
systems.\cite{Gros01} Since the presence of a substrate
alters these properties\cite{PLK+03,FGH+04} detailed investigations of
the substrate-adsorbate interaction\cite{VMG03} are required. 

In this paper we present results from transmission electron microscopy
(TEM) investigations\cite{Reim89,LL02} concerning the time- and
temperature-dependent behavior of Au clusters deposited by laser
ablation technique on amorphous-carbon (a-C) films. The reasons to
choose an amorphous substrate are threefold. Firstly, the
non-crystalline structure of the substrate allows for the
visualization of the crystal structure of the adsorbed clusters in
high resolution TEM (HRTEM) experiments. Secondly, amorphous carbon is
mechanically much more stable than highly oriented pyrolytic graphite
(HOPG) and consequently much more suited for technical
applications. Thirdly, as opposed to the conductor HOPG, amorphous
carbon is semi-conducting \cite{Robe86,RFG04} and as such is
expected to be less influential on the electronic properties of the
adsorbate. Furthermore, our observations were made on samples that
were stored under normal atmosphere for several months enhancing their
relevance for possible technical applications.  

\begin{figure}
     \centerline{
       \includegraphics*[width=0.45\textwidth,clip=true]
       {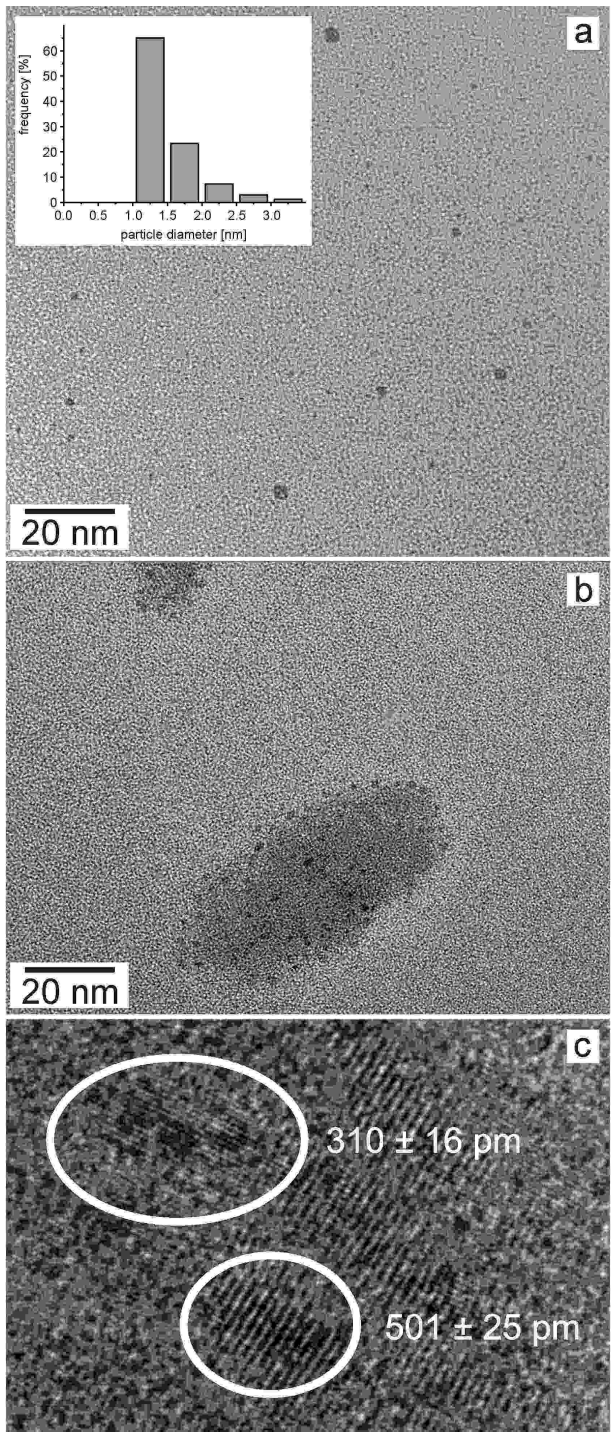}}
   \vspace{3ex}
   \caption{\label{expfig1}\sl 
     TEM images of Au clusters deposited on an a-C
     substrate by laser ablation (3000 shots): (a) sample one day after
     its preparation. The inset shows the histogram of the
     size-distribution of the clusters. Clusters smaller than 1nm in
     diameter cannot be resolved unambiguously within the TEM
     images. (b) Typical island of higher contrast formed on a sample
     four months after its preparation. (c) Detail of the island
     displayed in panel (b). The areas circled show regions exhibiting
     crystalline structures with differently spaced lattice planes. 
   }
   \end{figure}

The central observation reported in this paper is the formation of
islands on a time scale of several months after the deposition
of the Au adsorbate. Typical TEM images of the samples are shown in
Fig.\ \ref{expfig1}. Panel (a) is a sample one day after its
preparation. The inset shows the size distribution of the
clusters. Clusters smaller than 1 nm in diameter cannot be resolved
unambiguously because of their low contrast on the amorphous
substrate. Panel (b) shows a typical island formed after the sample
has been aged for four months. The reproducibility of the results is
assured since eight samples of different coverage prepared on different
days stored in individual sealed containers show the same phenomenon.

While the formation of small islands on short time scales has
been observed previously for larger size-selected clusters for Co on
microgrid substrates,\cite{PKW+01} for Au on amorphous
carbon,\cite{EHD02} as well as for Ag on graphite,\cite{CPP03} the new
features here are dark areas typically covering 25$\times$70 nm$^2$
underlying the larger clusters. As described in this paper in detail
we were able to identify these areas as islands consisting of Au films
of roughly 4 to 8 monolayers thickness with immersed larger Au
clusters. The Au films show areas of crystalline structure with
essentially two different lattice plane orientations as shown in the
circled areas in Fig.\ \ref{expfig1}(c). A close analysis of the
properties of the Au films including a holographic determination of
its thickness is given is Sec.\ \ref{sectiondynamics}. We arrive at
the conclusion that the islands are formed by coalescing small
clusters with diameters of $< 2$ nm.

Since the observed island formation was unexpected on the amorphous
substrate, no monitoring of the dynamics of the formation of the
islands---as desirable---was done. Awaiting the preparation of new
samples, which are needed for the observation of the formation
process, the present paper aims to give a state-of-the-art analysis of
the system at hand. Based on these findings the time-analysis will be
performed once new samples are available.

\subsection*{Outline of the paper}

In Sec.\ \ref{sectionmethods} we describe the experimental [sample
preparation and TEM] as well as theoretical [density functional theory
(DFT) and Monte-Carlo (MC) simulations] methods applied. 

Since the main trade-off of the amorphous substrate is its less well
defined surface we devote Sec.\ \ref{sectionsurface} to develop a
model of the carbon film that is consistent with the experimental
observations. To this end holographic images of the substrate film are
analyzed (Sec.\ \ref{SubstrateHolography}). 
First principle DFT calculations allow for the determination of the
substrate-Au interaction (Sec.\ \ref{SubstrateDFT}).  

In Sec.\ \ref{sectiondynamics} the observed islands are analyzed in
detail. Their thickness is determined holographically (Sec.\
\ref{sectionFilmChar}) and their stability tested in a heating
experiment (Sec.\ \ref{sectionTemp}). 

In Sec.\ \ref{sectionMCfilm} the formation of the islands and the shape
of the Au clusters on their surface is modeled with MC simulations
followed by the conclusion Sec.\ \ref{sectionConclusions}.

\section{Methods}\label{sectionmethods}

\subsection{Experimental: substrate and sample
  preparation}\label{sectionsamples}

The commercial a-C substrate films were produced by evaporation in a
carbon arc by Arizona Carbon Foils and distributed by Plano GmbH as
type S160. The films are mounted on a 200 nm mesh Cu grid for
support. The film thickness is given by the manufacturer as $d_{\rm
  subs} = 10 - 12.5$ nm with a density of $\rho_{\rm subs} \approx
2.0$ g/cm$^3$ corresponding to a mass of $2 - 2.5 \mu {\rm g}/{\rm
  cm}^2$. The density of the substrate is closer to that of graphite
($\rho_{\rm graph} = 2.267$ g/cm$^3$) than to that of diamond
($\rho_{\rm dia} = 3.515$ g/cm$^3$) suggesting a structure that
contains regions with trivalent coordination.\cite{Robe86} The
similarities in the electronic structure between a-C and graphite are
supported by electron energy loss spectroscopy (EELS)
measurements,\cite{RFG04} which show spectra that differ significantly
from those of diamond. Amorphous-carbon films produced with similar
methods and of similar density have been found to be
semi-conducting.\cite{Robe86,TL04}

Note that, since $\rho_{\rm subs} < \rho_{\rm graph}$ and since in the
amorphous material some tetravalent C with higher local density must
be present, there must exist areas with very low density or even voids
for compensation. This assumption is supported by the results presented
in Sec.\ \ref{sectionsurface}.

The Au clusters were collected from the primary beam of a laser
vaporization cluster source, which has been described elsewhere in
detail.\cite{WWVK04,WGGK02} 
 In brief, the laser vaporization cluster
 source is a variant of the  Smalley-deHeer-type \cite{MdH90,DDPS81}
 setup optimized by Heiz \cite{HVTS97} for high yield. The source is
 equipped with a rotating gold disc target with a diameter of 50 mm
 which is sealed with a Teflon gasket against the source block. A
 pulsed laser (Neodym-YAG, Continuum, 532 nm, 30 Hz repetition rate) is
 focused through a nozzle onto the target. A pulsed valve (General
 Valve, 5 bar backing pressure of He) which is synchronized with the
 laser quenches the evaporated atoms into clusters which expand through
 the nozzle and a skimmer into an oil diffusion pumped vacuum
 chamber at 10$^{-5}$ mbar. Au clusters (and atoms) are deposited
 without further mass selection onto a TEM grid placed in the primary
 beam in a distance of about 40 cm from the nozzle.

The size distribution of the Au particles is shown in the inset of
Fig.\ \ref{expfig1}(a). Since the only directed acceleration of the
clusters is the expansion into the vacuum of the cluster source the
impact energy can be considered small enough to avoid intercalation
of adsorbate and substrate.

The investigations were conducted on a batch of 8 samples prepared
on different days. Four samples were prepared with 300 ablation
shots, four with 3000 ablation shots. They were stored individually
under air in sealed containers. Experiments were carried out to
assure the absence of contamination on the samples (Sec.\
\ref{sectiondynamics}). All samples show the same island formation
assuring the reproducibility of the results.

\subsection{Experimental: transmission electron
  microscopy}\label{sectionTEM} 

The TEM was carried out with a Philips CM200 FEG/ST electron
microscope at an energy of 200 keV equipped with a Noran Ge detector
system for energy-dispersive X-ray analysis (EDX). A Gatan 652 Double
Tilt Heating Holder operated by a Gatan 901 SmartSet Hot Stage
Controller was used to perform the in situ annealing
experiments.

Transmission electron holography was carried out using a M\"ollenstedt
biprism installed in the selected-area aperture holder of the
microscope. The electrostatic potential of the biprism wire was close
to 150 V. The images recorded were analyzed using the phase shift of
the (000)-beam of the first hologram sideband. Data analysis was
performed using the DALI program package,\cite{RKR+96} which was
extended for the reconstruction of holograms. The details of the
reconstruction of holograms are outlined by Lichte and
Lehmann.\cite{LL02}

\begin{figure}
  \centerline{
    \includegraphics*[width=0.48\textwidth,clip=false]
    {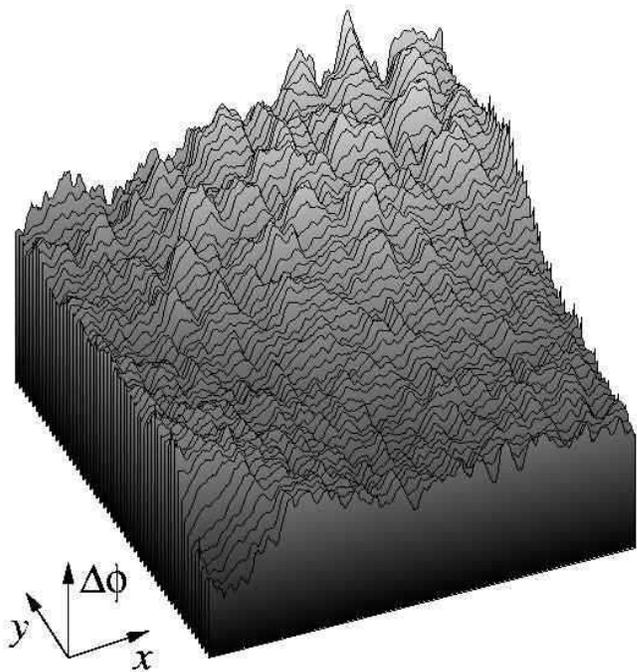}}
  \caption{\label{SampleHologram}\sl 
    Holographic phase shift $\Delta\phi$ of a segment of the a-C
    sub\-strate with part of an island as shown in Fig.\
    \protect\ref{expfig1}(b) in the far right corner. The image has
    the size of 28.2$\times$28.2 nm$^2$. The $z$-axis value is
    proportional to the observed phase shift. In the foreground the
    almost linear decrease of the a-C film thickness towards its edge
    is visible. (For alternative representations of the data see
    Figs.\ \protect\ref{holoscans} and \protect\ref{holo20diag}.)
  }
\end{figure}

Figure \ref{SampleHologram} shows the three-dimensional visualization
of the phase shift\cite{Phasesign} with respect to the vacuum
$\Delta\phi$ observed on a 28.2$\times$28.2 nm$^2$ surface segment of
a typical sample. In the far right corner a part of an island as shown
in Fig.\ \ref{expfig1}(b) is visible. In the front left corner the
linear decrease of the thickness of the substrate near the substrate
edge is visible. All holographic images were taken at substrate edges
in order include a section of the vacuum for
calibration.\cite{Reim89,LL02} A closer analysis of the data is given
in Sec.\ \ref{SubstrateHolography} concerning the substrate and in
Sec.\ \ref{sectionFilmChar} concerning the Au adsorbate.

The relation between the change in phase shift $\Delta \phi$ observed
in the holographic images and the corresponding thickness $d$ of the
observed object can be derived from scattering theory\cite{Reim89} and
is expressed via the standard formula 
\begin{equation}\label{thick}
  d = \frac{\Delta\phi}{V_0\ C_{\rm E}}\,,
\end{equation}
where the interaction constant depends on the acceleration voltage
resulting in $C_{\rm E} = 7.29 \times 10^{6}$ rad/(V m) in our case
and where $V_0$ denotes the effective internal potential of the
sample.  

For bulk gold the theoretical values\cite{Buhl59,Kell61} for $V_0$ are
$V_{\rm Au,theo} = 28\pm2$ V, while experimental values\cite{GC99} are
given as $V_{\rm Au,exp} = 22\pm1$ V. For small Au clusters on TiO$_2$
substrates a sharp increase of $V_0$ for clusters smaller than 4 nm
was observed with values of up to $V_{\rm Au,cluster} \sim 50$ V for
clusters smaller than 2 nm.\cite{IAO+03} We confirm these findings in
Sec.\ \ref{sectionFilmChar}.  

For the a-C substrate films the situation is less consistent. We
observe phase shifts of the substrate film with respect to the vacuum
of $\sim 1.9 - 2.8$ radians. Associating these phase shifts with the
numbers of the manufacturer for the film thickness of 9.5 to 14 nm we
obtain a potential of $V_{\rm subs} = 27.4$ V. 

On the other hand, the experimental value for a-C films
of the same density as ours has been given by Harscher and
Lichte\cite{HL98} as $V_{\rm HL} = 10.7$ V. This number compares 
satisfactorily with the values for graphite summarized by
S\'anchez and Ochando\cite{SO85} of $V_{\rm graph,theo}\sim 12 - 15$ V
(theory) and $V_{\rm graph,exp} \sim 11 - 13$ V (experiment) since the
ratio of the potentials $V_{\rm HL} / V_{\rm graph} \approx \rho_{\rm
  subs} / \rho_{\rm graph}$ corresponds roughly to that of the
densities, i.e., the inner potential scales roughly with the density
of the material. A similar argument holds for the values for diamond,
where  $V_{\rm dia,theo} \sim 16 - 23$ V (theory) and $V_{\rm dia,exp}
\sim 15 - 21$ V (experiment). 
 
The obvious discrepancy between the values of $V_{\rm subs}$ and
$V_{\rm HL}$ cannot be resolved here. Possible origins of the
discrepancies are (i) the underestimation of the substrate thickness
by the manufacturer\cite{RB98,AWWB03}  and (ii) an increase of the
internal potential for thin films similar to the increase of the
internal potential in small Au clusters.\cite{IAO+03} 
Note that the latter effects would not have been observed for the
thicker films with $d \ge 35$ nm in the investigations by Harscher and
Lichte.\cite{HL98} We will discuss the implications of the discrepancy
for the present work where appropriate.

\subsection{Theory: density functional theory calculations}

We have calculated the interaction between Au atoms and the carbon 
substrate using first principles DFT calculations in the generalized
gradient approximation (GGA).\cite{WP91} For all calculations we used
the highly accurate Projector Augmented  Wave method \cite{Bloe94} as
implemented in the VASP electronic structure program.\cite{KJ99}

For the carbon substrate we used both a graphite surface as well as a
model a-C surface consisting of 150 carbon atoms.\cite{DFTquote} The
Au-substrate interaction was calculated for a number of different
positions of an Au atom relative to the substrate surface. For each
position the Au-substrate interaction was mapped out by holding a Au
atom fixed at different heights above the surface while the carbon
atoms at and near the surface were allowed to fully relax. As a
reference point a Au atom at a distance of 9.5 \AA\ from the surface
was used (Sec.\ \ref{SubstrateDFT}). The resulting Au-substrate
potentials were averaged and fitted to a modified Lennard-Jones
potential and used as input in MC simulations described in the next
section.

\subsection{Theory: Monte-Carlo (MC) simulations}\label{sectionDefMC}

In order to develop a microscopic understanding of the dynamics of the
Au clusters on the substrate we simulate the system with the canonical
Monte-Carlo method. A standard Metropolis algorithm is employed
\cite{AT89,WDS+01,Wern05a} with an update after each random
displacement of an Au atom within an interval $[0,d_{\rm max}]$ in all
spatial dimensions. $d_{\rm max}$ is set to yield an MC acceptance
rate of 50 to 60 \%. The resulting temperature dependence is roughly
$d_{\rm max} \propto \sqrt{T}$. The boundary conditions are imposed by
a hard wall cube with linear dimension $L_x$, $L_y$, and $L_z$. 

The Au-Au interaction is modeled via the many-body Gupta
potential\cite{Gupt81} (GP):  
\begin{eqnarray}\label{Gupta}
V(\{r_{ij}\}) &=& \sum_{i}^N
  \sum^N_{j\neq i}A\ e^{-p(r_{ij}/r_0 - 1)}  
\nonumber\\&&\hspace{6ex}-\ 
  \sum_{i}^N
  \sqrt{\sum_{j\neq i} \xi^2\ e^{-2q(r_{ij}/r_0 - 1)}}\ .
\end{eqnarray}
The distances $r_{ij} = |\br_i - \br_j|$ are measured in units of the
bulk first-neighbor distance $r_0 = 2.885$ \AA. The indices $i,j \in
\{1,\ldots,N\}$ label the Au atoms at positions $\br_i$ and $\br_j$,
respectively. The parameterizations for Au as found in the
literature,\cite{CR93} i.e., $A = 0.2061$ eV, $\xi = 1.790$ eV, $p =
10.229$, and $q = 4.036$, has been determined to match the bulk
elastic constants and the surface contraction.

The Au-substrate interaction is modeled by two-particle interactions
in the form of generalized Lennard-Jones 6-12 and 3-6
potentials. Their derivation is discussed in detail in Sec.\
\ref{SubstrateDFT}.

Runs on the atomic level have been performed for up to 10$^{8}$
updates per atom for up to 600 atoms. Runs have been stopped when a
metastable configuration is obtained that does not evolve anymore on a
reasonably accessible time scale. Depending on the size of the system
metastable configurations are attained after a few minutes ($N = 55$)
or a couple of days (Au double layer formation with embedded clusters
with $N=600$). Runs where performed up to three weeks to assure that
the lifetime of the metastable configuration is at least an order of
magnitude larger than its formation time. Run times for particular
cases are given with the resutls in Sec.\ \ref{sectionMCfilm}. 

In order to simulate larger systems with many clusters we derive in
Sec.\ \ref{sectionClusterPot} an effective cluster-cluster
potential. Runs are then performed with a few hundred clusters to
illustrate the cluster diffusion and fusion process in Sec.\
\ref{sectionMCislands}.



\section{Substrate characterization}\label{sectionsurface}

The observed formation of the islands as shown in Fig.~\ref{expfig1}
implies that the substrate surface is free of strong pinning
centers. The same conclusion can be drawn from oval shapes of the
islands with smooth boundaries. Moreover, the pattern formation of the
Au film discussed in Sec.\ \ref{sectionIsland} [Fig.\
\ref{expfig1}(c)] strongly suggest the presence of smooth areas on the
substrate surface.

Since this observation is not ad hoc intuitive for an amorphous
substrate we investigated the substrate with off-axis holography
experiments. In order to obtain a qualitative understanding of the
underlying mechanisms of the island formation we determine the
substrate-adsorbate interactions via DFT calculations. The latter are
used as input parameters for MC simulations which reproduce the
experimental observations.

\subsection{Holography}\label{SubstrateHolography}

\begin{figure}
  \centerline{
    \includegraphics*[width=0.475\textwidth,clip=true]
    {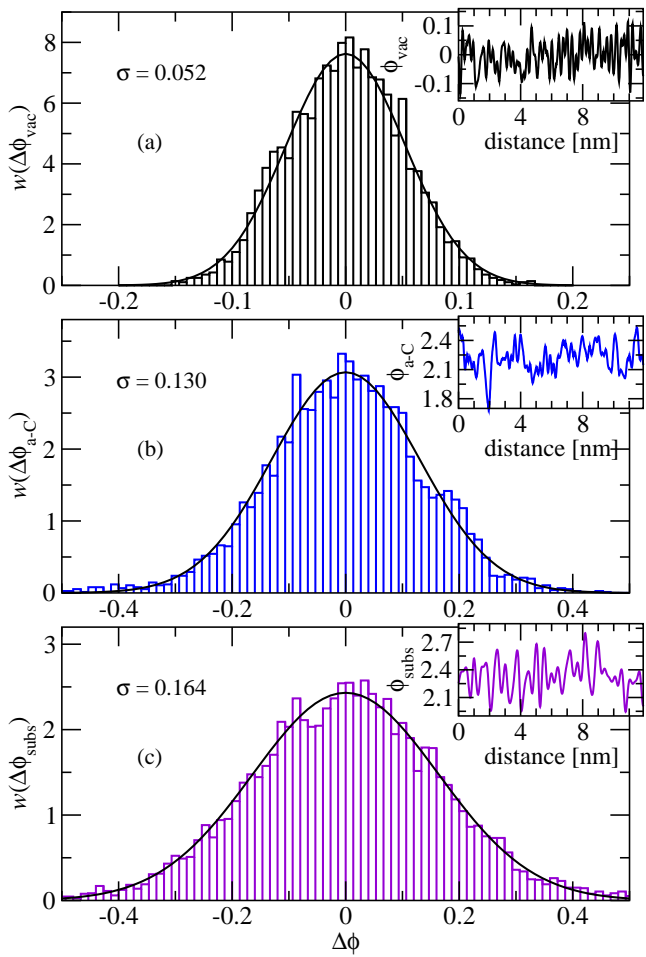}}
   \caption{\label{SubstrateHologram}\sl (Color online)
     Phase shift distributions as a measure of fluctuations of the
     phase. Panel (a) shows the result for the vacuum, panel
     (b) shows the substrate film without adsorbate, and panel (c)
     shows the distribution for the substrate film with Au adsorbate
     albeit at positions without islands. Please note the different
     scales of the $x$ axes. Histograms are obtained by binning and
     normalization of the shifted raw data, full lines are Gaussian
     fits to Eq.\ (\protect\ref{Gauss}). The insets show samples of
     the measured phase shifts. The data used to obtain the histograms
     amounts 5 to 10 times the date shown in the insets.}
   \end{figure}

When imaging vacuum with TEM off-axis holography the phase shift
observed is slightly fluctuating due to imperfections of the biprism
wire, aperture, and noise induced by the CCD camera.\cite{LL02} We
analyzed these vacuum fluctuations in order to distinguish them from
the sample signal. The inset in Fig.\ \ref{SubstrateHologram}(a) shows
a typical image of the vacuum phase shift $\phi_{\rm vac}$ after
subtraction of its mean value. The main graph of Fig.\
\ref{SubstrateHologram}(a) shows the normalized distribution $w_{\rm
  vac}(\Delta\phi_{\rm vac})$ of the vacuum fluctuations
$\Delta\phi_{\rm vac} = \phi_{\rm vac}$. The full line is a Gaussian
fit  
\begin{equation}\label{Gauss} 
w(\Delta\phi) = \frac{1}{\sqrt{2\pi}\sigma}\ 
\exp\left\{-\frac{(\Delta\phi)^2}{2\sigma^2}\right\} 
\end{equation}
from which we obtain a width of the vacuum fluctuations of $\sigma =
\sigma_{\rm vac} = 0.052$ rad.

The histogram in Fig.\ \ref{SubstrateHologram}(b) shows the
normalized phase shift distribution of the a-C substrate film {\em
  without} Au adsorbate. The Gaussian fit from Eq.~\ref{Gauss} as
shown by the full line has a width of $\sigma_{\rm a-C} = 0.13$
rad. Since $\sigma_{\rm a-C} > 2 \sigma_{\rm vac}$ the fluctuations
must reflect properties of the substrate. Using Eq.~\ref{thick} and
the value of $V_{\rm subs} = 27.4$ V the change in thickness
corresponds to $\Delta d_{\rm subs} = 2 \sigma_{\rm subs}/(C_{\rm
  E}V_{\rm subs}) = 1.3$ nm, for $V_{\rm HL} = 10.7$ V even $\Delta
d_{\rm subs} = 3.3$ nm. The graph in the inset Fig.\
\ref{SubstrateHologram}(b) reveals that phase fluctuations of
comparable depth occur on length scales of 0.3 nm. Such narrow
intrusions with a depth of up to 1/5 of the substrate thickness would
make the substrate very unstable. Moreover, the edges of the
intrusions would inevitably lead to strong pinning centers similar to
those of nanopits in HOPG,\cite{HBB+97,Hoev01} which are inconsistent
with the observed island formation. 
 
Since the intrusions are only 2 to 4 C-C bond lengths wide, it appears
likely that they are capped in the production process thus stabilizing
the structure. The intrusions are then left as voids in the film. The
effective potential is strongly reduced at the position of the voids
leading to the observed phase modulation. At the same time the cap
yields a rather smooth surface consistent with the observed island
formation. The C atoms surrounding the voids have less nearest
neighbor atoms,\cite{DD98} which is consistent with the measured EELS     
spectra\cite{RFG04} indicating largely trivalent coordination. In the
absence of methods for a more precise determination of the local
structure of the substrate we use this scenario as a working
hypothesis.

Finally, the histogram in Fig.\ \ref{SubstrateHologram}(c) shows the
normalized phase shift distribution of the a-C substrate film {\em
  with} the Au adsorbate, albeit at positions without any of the
observed islands. (The latter are discussed in Sec.\
\ref{sectionFilmChar}). The Gaussian fit from Eq.~\ref{Gauss} as 
shown by the full line has a width of $\sigma_{\rm subs} = 0.164$
rad which is 25\% larger than $\sigma_{\rm a-C}$. The difference is
readily interpreted as induced by small Au clusters pinned at small
intrusions of the substrate. Since the pinned clusters are smaller
than the TEM resolution limit their diameters must be smaller 1 nm or
$N<50$ atoms.\cite{Urba98}

\subsection{Surface adsorbate interaction: DFT
  calculations}\label{SubstrateDFT} 

A quantitative description of Au/a-C substrate interaction is
needed for the MC simulations described in Sec.\ \ref{sectionMCfilm}.
To this end we studied the interaction of individual Au atoms with
graphite and a model a-C surface using ab-initio DFT
calculations.

\begin{figure}
     \includegraphics*[width=0.475\textwidth,clip=true]
     {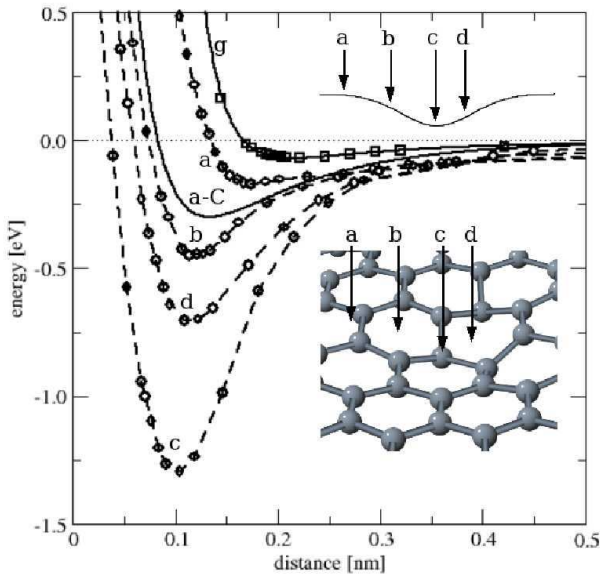}
     \caption{\label{Au-substrate-potentials}\sl (Color online)
       Au/graphite (squares) and Au/a-C substrate potentials
       (circles) as calculated within the DFT approach.
       The inset shows the corrugated surface layer of
       the a-C model with the locations of the calculated
       potentials (a)-(d) indicated by arrows.
       The fitted potentials used in the MC simulations are shown as
       solid lines for graphite (g) and the a-C model substrate.
       Dashed lines are guides to the eye.}
   \end{figure}

For the calculation of the Au/graphite-surface interaction we used a 
supercell consisting of 2x2 graphite unit cells in the $x$-$y$ plane
and four carbon layers in the $z$-direction. A large vaccuum distance of
19 \AA\ was used.\cite{DFTquote} Calculations were performed for Au in the
top, bridge, and hollow sites. Using as reference a Au atom at the
center of the vacuum we calculated binding energies of 0.065 eV, 0.062
eV and 0.041 eV, respectively. The Au-graphite potential for Au in the
top-site as a function of Au atom surface distance is shown by line
``g'' in Fig.\ \ref{Au-substrate-potentials}. The result of the DFT
calculations was fitted with a modified Lennard-Jones potential 
\begin{equation}\label{LJfitgraphite}
V_{\rm LJ,graph} = V_{{\rm LJ},0}\left( \frac{0.31}{(z/r_0 - 0.1)^6} - 
  \frac{1.11}{(z/r_0 - 0.1)^3} \right),
\end{equation}
where $V_{{\rm LJ},0}$ is the binding energy and $r_0 = 2.885$ {\AA}.


For a more realistic description of the a-C substrate we have
constructed a model a-C surface. To ensure that sp$^2$ bonding will be
dominant\cite{RFG04} and to obtain a simple model consistent with the
observed Au-film formation we use graphite as our starting point. We
use a supercell consisting of 4x4 graphite unit cells in the x-y plane
and four carbon layers in the z direction. Additional carbon atoms were
added in localized regions but otherwise randomly inbetween the
graphite planes so as to create voids consistent with the observed
holographic phase fluctuations.\cite{DFTquote}

The resulting supercell of 150 carbon atoms was relaxed into a local
minimum. Two main effects of the relaxation can be observed: The
carbon structure expanded strongly in the z-direction and the graphite
layers buckled due to the uneven distribution of interstitial carbon
atoms. The density of the final a-C model structure is $\rho\sim2.0$
g/cm$^3$ in good agreement with the density of the a-C substrate as
given by the manufacturer. The surface layer of the final structure is
smooth but has a corrugation of $d = 0.9$ \AA\ over a distance of a
few C-C bondlengths (Fig.\ \ref{Au-substrate-potentials}).    

We have calculated the Au/a-C substrate potential at several
positions roughly along a line through a depression in the
surface. Since for the graphite surface the Au on top position is the
preferred bonding site, only Au on top sites were considered in the a-C
case. High lying C surface atoms retain their graphite character and
are predominantly 3-fold coordinated. Accordingly we find the binding
energy of an Au atom is still comparable to the binding energy on
graphite (case a in Fig.\ \ref{Au-substrate-potentials}).  C atoms
lying in a depression of the surface have additional bonds to the
nearby C atoms below and are mostly fourfold coordinated. The calculated
binding energies of an Au atom to C sites lying in such a depression
are considerably larger as compared to the graphite surface (cases b,d
in Fig.\ \ref{Au-substrate-potentials}). An Au atom in the center of
the depression can bind to several C atoms and the binding energy is
smaller but of the same order as the Au-Au binding energy (case c in
Fig.\ \ref{Au-substrate-potentials}).

To model the binding energy of Au atoms in clusters with a contact
area larger than the surface intrusions the resulting Au/a-C substrate
potentials were averaged using approximate relative surface areas as
weights and fitted to a modified Lennard-Jones potential
\begin{equation}\label{LJfitmean}
V_{\rm LJ,mean} = V_{{\rm LJ},0} \left( \frac{297}{(z/r_0-1.2)^{12}} - 
  \frac{34.5}{(z/r_0-1.2)^6} \right).
\end{equation}
where $V_{{\rm LJ},0}$ is the binding energy and $r_0 = 2.885$\ {\AA}. 
Using weights of c:d:b:a=1:4:9:16 results in an average binding energy of
$V_{{\rm LJ},0}=0.34$eV. The MC simulations in Sec.\
\ref{sectionMCfilm} do not depend qualitatively and only 
little quantitatively on the specific value in a range of $0.2 <
V_{{\rm LJ},0}/{\rm eV} < 0.4$ and hence an average value of 
$V_{{\rm LJ},0} = 0.3$eV was adopted. 

The value of $V_{{\rm LJ},0} \sim 0.3$ eV is only appropriate for
sufficiently large clusters. Small clusters with contact areas smaller 
than the intrusion size have larger weight of the strongly binding
intrusion center c and are easily pinned consistent with the 
observations discussed in Sec.\ \ref{SubstrateHolography}.


\section{Long time-scale dynamics}\label{sectiondynamics} 

The long time-scale rearrangement of heavy atoms on a-C substrates has
been observed previously.\cite{Utla80} Here we present a detailed
investigation of pattern formation of Au on a-C substrates. Figure
\ref{expfig1}(a) shows a typical TEM image of a sample prepared 
by deposition of 3000 shots of Au clusters on a 10 nm a-C 
film. The image displayed was recorded one day after the preparation
of the sample. As expected, one observes a statistical particle
distribution. The histogram included shows the size distribution. Au
clusters smaller than 1 nm in diameter could not be discerned from the
10 nm carbon substrate background owing to their low
contrast. However, ion mobility measurements have
revealed\cite{GWF+02} that clusters in this range of sizes are widely
present in cluster ion beams. This suggests that a significant number
of small clusters is also present on the substrate, albeit unresolved
in the TEM.

\subsection{Island formation}\label{sectionIsland}

After the sample is kept at room temperature and in absence of inert
conditions for four months, TEM studies yield images as displayed in
Fig.\ \ref{expfig1}(b). The previously statistically distributed 
clusters are now collected in agglomerates referred to as islands. The 
vast majority of these islands is of similar size with typical areas
of $A \sim 25 \times 70$ nm$^2$.

In all islands an underlying area of higher contrast is observed,
which is bordered by the outer clusters of the islands. Using
EDX-detection no elements except for Au and C are found in these
islands. The binary alloy phase diagram of Au and C excludes the
formation of Au-C alloys under the given conditions\cite{McLe69} and
we conclude that the C signature stems from the substrate. The
analysis given in Secs.\ \ref{sectionFilmChar}, \ref{sectionThick},
and \ref{sectionTemp} consistently suggest that the islands are formed
of 4 to 8 monolayers of Au on top of the substrate. This conclusion is
supported by the MC simulations of the formation process of such an Au
film in Sec.\ \ref{sectionMCfilm}.

HRTEM images of the islands as depicted in Fig.\ \ref{expfig1}(c) show 
regions, in which crystalline patterns occur with lattice spacings
determined as 501 pm and 310 pm within an estimated error of $\pm
5$\%. The angles between these two spacings differ up to 6$^\circ$
from a rectangular configuration. It cannot be excluded that much
larger areas of the Au films are crystallized forming such patterns
because a slight corrugation of the substrate may obscure its
detection with HRTEM.

While the exact determination of the origin of the aforementioned
patterns from the present data is not possible, it can be speculated
that the Au film exhibits a similar reconstruction as has been
observed for Au(111) surfaces by scanning tunneling microscopy (STM)
measurements,\cite{HCZC95,HCR+87} where superstructures with a similar 
modulation\cite{HKS+81} ($\sqrt{3} \times {\rm lattice\ constant}
\approx 500$ pm) have been observed. The presence of Au(111)
surfaces is also consistent with the MC simulations (Sec.\
\ref{sectionMCfilm}). Moreover, reconstruction pattern have been
observed for small islands of Au monolayers on HOPG in ultra high
vacuum (UHV) by STM measurements,\cite{GSC88} but a direct comparison
is difficult because of the larger film thickness in our samples
(Sec.\ \ref{sectionFilmChar}).

We conclude that the higher contrast areas underlying the islands
consist of a few monolayers of Au that form crystalline structures on
at least parts of the substrate surface, which in turn must be
sufficiently smooth.

\subsection{Au-film characterization}\label{sectionFilmChar}

\begin{figure}
     \centerline{
       \includegraphics*[width=0.475\textwidth,clip=true]
       {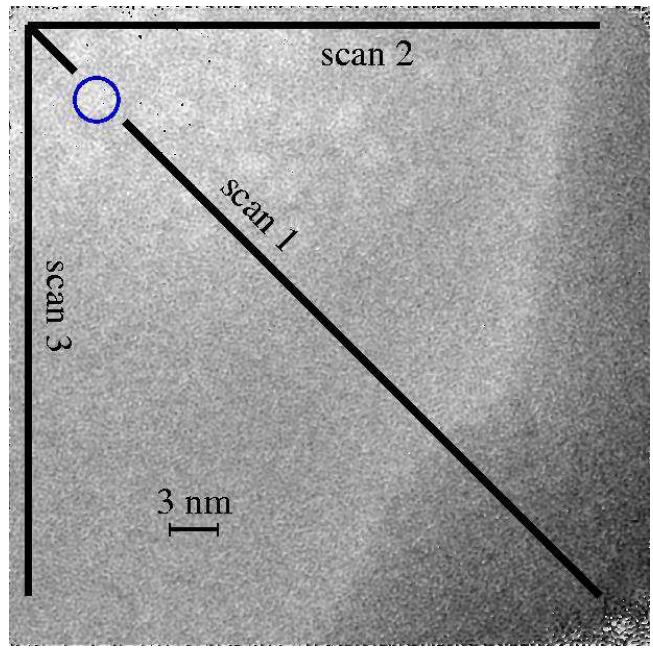}}
     \caption{\label{holoscans}\sl (Color online)
        Holographic phase reconstruction image from the same segment
        as shown in Fig.~\protect\ref{SampleHologram} as an intensity
        plot. Lighter shades of gray correspond to larger phase shifts
        with respect to the vacuum. Lines show scans that
        underwent closer investigation, circles indicate the positions
        of selected adsorbed clusters.}
   \end{figure}

In order to get information about the thickness of the observed films,
we performed off-axis holography experiments. A three dimensional
visualization of a segment of the substrate including a part of an Au
island has been shown in Fig.\ \ref{SampleHologram} in Sec.\
\ref{sectionTEM}. Figure \ref{holoscans} shows the same segment as an
intensity plot, where lighter shades of gray correspond to larger
phase shifts. The thick lines correspond to different scans
investigated in detail. The circled areas indicate the positions
of clusters adsorbed on the Au films that have been studied closely.

\begin{figure}
     \centerline{
       \includegraphics*[width=0.475\textwidth,clip=true]
       {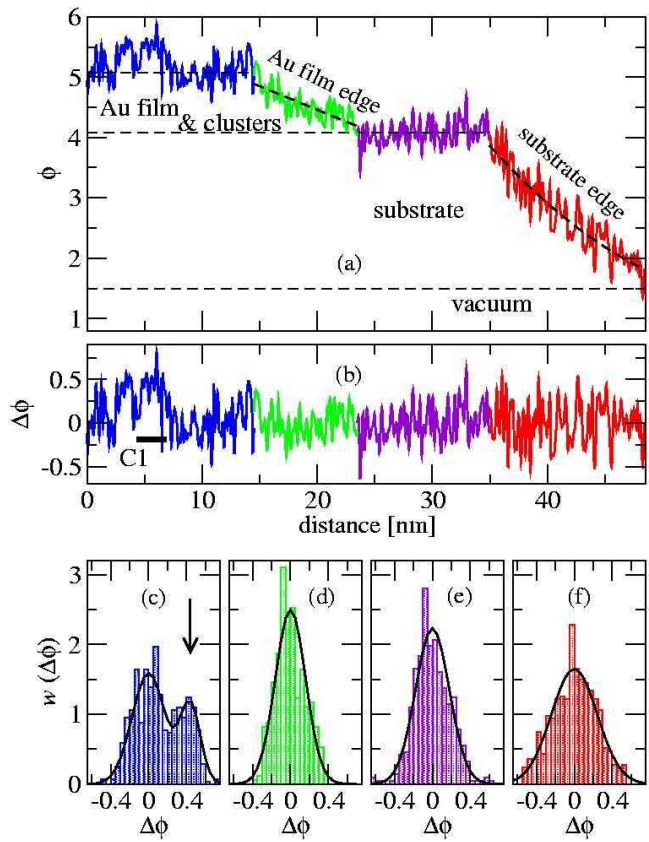}}
     \caption{\label{holo20diag}\sl (Color online)
       Details of scan 1 in Fig.~\protect\ref{holoscans}. Panel (a)
       shows the raw data of the phase shifts together with fitted
       values (dashed lines, see text), which average the
       fluctuations. Panel (b) shows the fluctuations $\Delta\phi$
       obtained from the raw data after subtraction of the
       fits. Panels (c), (d), (e), and (f) show the binned data of the
       phase fluctuations form the Au film, the Au-film edge, the
       substrate, and the substrate edge, respectively. Solid lines
       are fits from Eqs.\ (\protect\ref{doubleG}) and
       (\protect\ref{Gauss}).}  
   \end{figure}

The results for the three scans are qualitatively equivalent and for
simplicity we focus the presentation on scan 1. Figure
\ref{holo20diag}(a) shows the corresponding phase profile. Four
regions are clearly distinguishable from right to left: the rise of
the substrate edge, the plateau of the substrate film, the rise of the
Au-film edge, and finally the Au-film. The dashed lines 
show the profile after averaging the phase fluctuations. The vacuum
level was determined in a larger area a bit further away from the
sample (not shown). For the substrate edge the dashed line is a
quadratic fit, for the Au-film edge it is a linear fit. The substrate
plateau is given by a horizontal line at the average value of phase in
that region. Finally, the Au-film base line is obtained by adapting
the average value of the fluctuation phase without the adsorbed
clusters as outlined closer below.

Panel (b) of Fig.\ \ref{holo20diag} shows the fluctuation of the data
after subtraction of the dashed lines shown in panel (a). The black
bar labelled ``C1'' indicates the position of the corresponding
cluster as shown in Fig. \ref{holoscans}.

The lower panels of Fig.\ \ref{holo20diag} show the histograms
of the phase distribution in panel (b) for the Au film (c), the
Au-film edge (d), the substrate (e), and the substrate edge (f). The
full lines are Gaussian fits from Eq.\ (\ref{Gauss}) except for panel
(c), where a double Gaussian of the form 
\begin{equation}\label{doubleG}
w_{\rm tot}(\Delta\phi) = a_{\rm Au}\ w_{\rm film}(\Delta\phi) 
+ (1-a_{\rm Au})\ w_{\rm clus}(\Delta\phi)
\end{equation}
with Gaussian contributions from the film $w_{\rm film}$ and the
adsorbed clusters $w_{\rm clus}$ was used. The width of the curve for
the substrate of $\sigma_{\rm subs} = 0.18$ rad is consistent with the
averaged value obtained for a number of samples $\sigma_{\rm subs} =
0.164$ as discussed in Sec.\ \ref{SubstrateHolography}. The difference
may be attributed to poorer statistics of the relatively small sample
size in Fig. \ref{holoscans} and possibly to an increased number of
pinned Au clusters near the islands.

The substrate edge is expected to show growth process dependent steps
which are yet too small to be resolved in the noise of the density and
vacuum fluctuations. Since a simple polynomial fit cannot account for
this non-monotonous increase in thickness at the substrate edge, a
broader distribution of the phase fluctuations is expected in that
region. Indeed, the distribution for the substrate edge $\sigma_{\rm
  subs-edge} = 0.24$ [Fig. \ref{holoscans}(f)] is 30\% broader than
for the substrate plateau. Here a quadratic fit was applied to the
substrate edge but linear or cubic fits give very similar results.

The width of the phase distribution of the Au-film edge $\sigma_{\rm
  Au-edge} = 0.16$ in panel (d) is close to the average value of the
substrate, no smoothing of the surface corrugation is measurable. This
result is consistent with the previously discussed interpretation
(Sec.\ \ref{SubstrateHolography}), that the phase fluctuations of the 
substrate stem predominantly from intrinsic spatial density
inhomogeneities.

Finally, the bimodal distribution of the Au film in panel (c) of Fig.\
\ref{holo20diag} reflects the presence of adsorbed clusters on the
substrate with an average height in this segment resulting in an
additional phase shift of $\Delta\phi_{\rm clus} \approx 0.43$ rad as
indicated by the arrow. The width of the film part of the distribution
$\sigma_{\rm film} = 0.175$ rad is comparable to that of the substrate
and stems from the intrinsic spatial density inhomogeneities of the
latter. For completeness we note the values of the cluster
distribution width $\sigma_{\rm clus} = 0.11$ rad and the weighing
factor $a_{\rm Au} = 0.695$.

\subsection{Au-film thickness}\label{sectionThick}

In principle the thickness of the Au film can be obtained from Eq.\
(\ref{thick}). As discussed in Sec.\ \ref{sectionTEM} the parameter of
the mean inner potential $V_0$ is not well defined as it appears to
become thickness dependent for thin samples. To be specific, Ichikawa
and coworkers\cite{IAO+03} found for Au clusters on a TiO$_2$
substrate with diameters smaller than 4 nm an increase of $V_0$ of up
to a factor of 3. In order to have a reference point for our system we
have investigated the observed phase shifts of the cluster labelled
with ``C2'' in Fig.\ \ref{holoscans} and compared the results with the
shape of a cluster of the same size obtained in the MC simulations
described in Sec.\ \ref{sectionAdsorbed}. 

\begin{figure}
     \centerline{
       \includegraphics*[width=0.475\textwidth,clip=true]
       {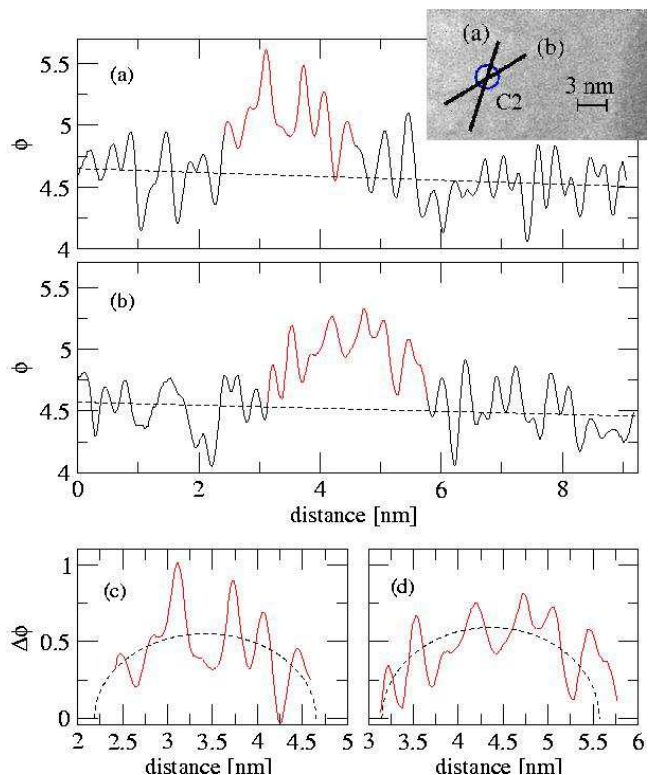}}
     \caption{\label{C2xgraph}\sl (Color online)
       Phase shift for two scans (a) and (b) through cluster ``C2'' in
       Fig.\ \protect\ref{holoscans} along the directions indicated in
       the inset. Dashed lines are linear fits to the background.
       Panels (c) and (d) show the data of the cluster from panels (a)
       and (b), respectively, after background subtraction. Dashed
       lines in (c) and (d) are elliptical fits (see text).} 
   \end{figure}

Panels (a) and (b) of Fig.\ \ref{C2xgraph} shows the resulting phase
shift profiles as obtained along the scans indicated in the
inset. Dashed lines are linear fits. Panels (c) and (d) show the phase
profiles of the cluster after subtraction of the linear contributions
together with fitted ellipses (dashed lines). As a result of the
strong fluctuations the fits are not unique. A width of $b_{\rm clus} =
2.3\pm0.2$ nm can be extracted while the phase shift is determined as
$\Delta\phi_{\rm clus} = 0.5\pm0.02$ rad. Comparing these values to
the aspect ratio of $h_{\rm clus}/b_{\rm clus} \approx 0.54$ as
obtained in Sec.\ \ref{sectionAdsorbed} a height of $h_{\rm clus} =
1.2\pm0.1$ nm is expected. Using Eq.\ \ref{thick} we obtain a mean
inner potential of $V_0 = 57\pm5$ V consistent with the results from
Ichikawa {\it et al.}\cite{IAO+03}

We have consequently confirmed the thickness dependence of the mean
inner potential of Au clusters smaller than 4 nm in
height. Unfortunately, when approximating the apparent strong increase
of the inner potential with decreasing thickness\cite{IAO+03} in
leading order as $V_0\sim d^{-1}$, the phase shift $\Delta\phi \propto
V_0\ d$ becomes {\em independent} of the thickness in leading
order. From the phase shift induced by the Au film as measured in
Fig.\ \ref{holo20diag}(a) of $\Delta \phi_{\rm film} = \phi_{\rm film}
- \phi_{\rm subs} \approx 1$ rad we can thus only determine an upper
bound for the thickness of the Au film of $d_{\rm film} < 7$ nm. This
upper bound is obtained from Eq.\ (\ref{thick}) using the bulk
value\cite{GC99} $V_{\rm Au,exp} \approx 22$ V, which is a lower bound
for the inner potential. A more precise determination requires the
detailed knowledge of the sub-leading contributions to $V_0(d)$.

The intensity profiles in Figs.\ \ref{expfig1}(b) and \ref{expfig2}(a)
suggest that the film is thinner than the majority of the adsorbed
clusters. This leads to an estimate of $d_{\rm film} \sim 4-8$
monolayers. The observation discussed in Sec.\ \ref{sectionMCfilm}
that clusters of size $N < 300$ (or $d_{\rm cluster} < 2$ nm) are not 
stable enough to retain their structure on the substrate at room
temperature and should consequently largely contribute to the film
formation is consistent with these numbers.

\subsection{Heating effects}\label{sectionTemp}

\begin{figure}
   \epsfxsize=0.48\textwidth
   \centerline{\epsffile{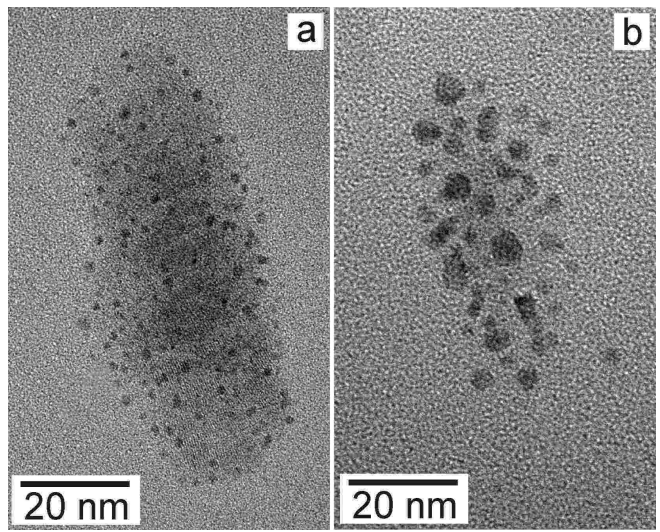}}
   \vspace{3ex}
   \caption{\label{expfig2}\sl 
     TEM images of two typical  islands  (a) before and (b) after {\em
       in situ} annealing at 373 K for 2 h. 
   }
   \end{figure}

The effects of heating on the shape of metal clusters has been
studied early on.\cite{DMH81,SPWB86} In order to obtain information
about the stability of the islands we performed experiments heating
the sample in situ. The sample was kept at 373 K for two hours. Figure
\ref{expfig2}(a) displays the characteristics of a typical island
before heating, whereas another island, which was not exposed to the
electron beam during annealing, is shown in Fig.\
\ref{expfig2}(b). Comparing these two images, it is obvious that as a
consequence of the heating (i) the amount of Au forming the film is 
diminished, (ii) the number of particles is reduced and (iii) the
sizes of the remaining particles have increased.

The effects (ii) and (iii) are typical characteristics of
Ostwald-ripening processes.\cite{Ostw00} The islands are metastable
with an activation energy low enough that the applied heating of 100 K
above room temperature is sufficient to induce the transition on a
time scale of a few hours. A more elaborate investigation of the
dynamical behavior of the Au islands at elevated temperatures requires
more experimental data and will be presented in a subsequent
publication.


\section{Monte-Carlo simulations}\label{sectionMCfilm}

In order to obtain a more microscopical understanding of the formation
of the Au islands---which cannot be resolved experimentally---we
performed MC simulations based on the Au-substrate interaction
elaborated in Sec.\ \ref{sectionsurface} and using the Gupta many-body
effective potentials Eq.\ (\ref{Gupta}) as described in Sec.\
\ref{sectionDefMC}.

\subsection{Cluster fusion and effective cluster
  potential}\label{sectionClusterPot}

As discussed in Sec.\ \ref{sectionIsland} the observed Au film is
likely to be composed of fused small Au clusters, which are present at
the time of deposition but only unambiguously resolved in TEM for
diameters larger than 1 nm (c.f.\ Fig.\ \ref{expfig1}). The fusion
process of adsorbed clusters has been observed experimentally for Co
clusters\cite{PKW+01} of diameter 8.5 nm as well as for
clusters\cite{EHD02} of Au$_{5000}$ and and has been modeled for free
clusters with MD methods using embedded atom\cite{LJB97} and glue
potentials\cite{ABPG04} for Au$_{225}$ to Au$_{3805}$. The objective
here is to obtain an effective inter-cluster potential in the
framework of the better suited many body Gupta potentials Eq.\
(\ref{Gupta}), which we will use in Sec.\ \ref{sectionMCislands} to
simulate the island formation.

DFT results\cite{HCSR97} show that the icosahedral configuration has a
lower cohesive energy than the octahedral or cuboctahedral structures
for the atomic closed shell cluster sizes $N=13,55,147$. On the other
hand, the Gupta potentials for Au exhibit a large number of local
minima with amorphous structures and energies of $\Delta E_{55} \sim
0.01$ eV lower than the icosahedral configuration.\cite{GMB+98} Since
$\Delta E_{55} \sim 0.01\ {\rm eV} < k_{\rm B} T_{\rm room} = 0.0256$
eV, we use icosahedral configurations for the investigations presented
here because different isomers are realized through thermal
fluctuations at room temperature.\cite{DW98b} Note that only
quantitative details depend on the specific structure while our
results are qualitatively quite general.

\begin{figure}
     \centerline{
       \includegraphics*[width=0.475\textwidth,clip=true]
       {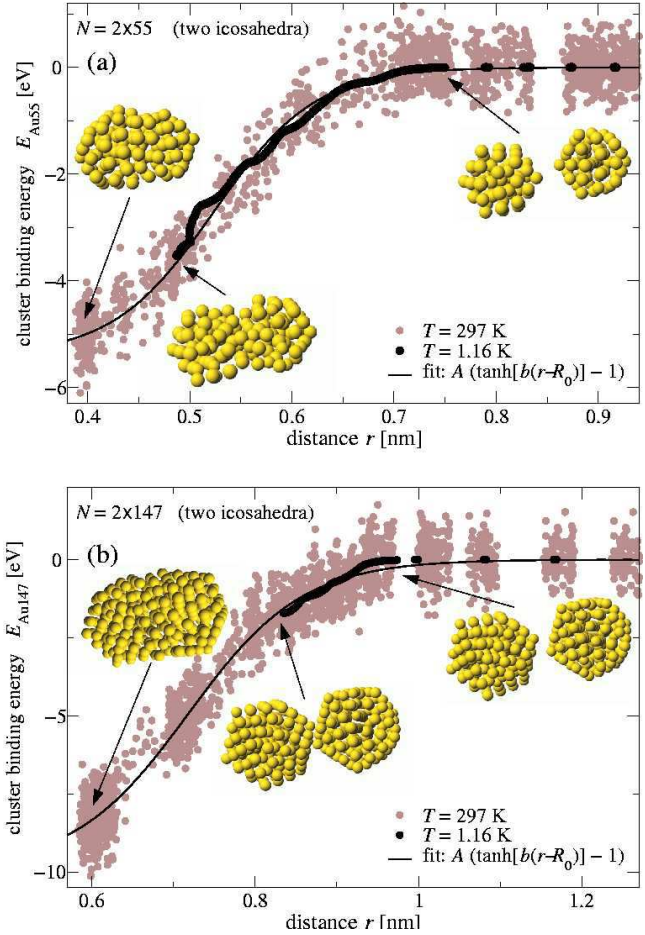}}
     \caption{\label{clusterpot}\sl (Color online)
       Fusion of Au$_N$ clusters for (a) $N=55$ and (b) $N=147$. Gray
       and black circles are obtained in MC annealing runs at $k_{\rm
       B} T = 0.0256$ eV ($T = 297$ K) and $k_{\rm B} T = 0.0001$ eV
       ($T = 1.16$ K), respectively. The full lines are fits for the
       effective inter-cluster cohesive potentials Eq.\
       (\protect\ref{ClustercohesiveE}). At low temperatures the
       fusion process comes to an early halt because the energy barrier 
       involved in the atomic rearrangement cannot be overcome.}  
   \end{figure}

Figure \ref{clusterpot} shows the inter-cluster cohesive energy as a
function of the distance of the center of mass points $r$ of the two
clusters. Panels (a) and (b) were obtained for two Au$_{55}$ and two
Au$_{147}$ clusters, respectively. The gray points are obtained in
free MC annealing runs at room temperature ($k_{\rm B} T = 0.0256$ eV)
while the black circles were obtained at $k_{\rm B} T = 0.0001$ eV ($T
= 1.16$ K). The scattering of the gray data points reflects the
thermal activation of the clusters. At low temperatures the fusion 
process is halted at larger distances than at room temperature because
the potential energy barrier involved in the reconstruction cannot be
overcome.

The full lines in Figs.\ \ref{clusterpot}(a) and (b) are fits
representing the effective inter-cluster cohesive energy and have the
form
\begin{equation}\label{ClustercohesiveE}
E_{{\rm Au}N} = A_N \left\{\tanh\left[b_N(r-R_{0,N})\right] -1\right\}\,.
\end{equation}
The parameters extracted are $A_{55} = 2.702$ eV, $b_{55} = 10.08$
nm$^{-1}$, $R_{0,55} = 0.524$ nm and $A_{147} = 4.941$ eV, $b_{147} =
7.149$ nm$^{-1}$, $R_{0,147} = 0.718$ nm. The fusion process stops at
a temperature dependent minimal distance $R_{{\rm min},N}(T)$ which
can be modeled in the potential as a step function with a high
repulsive value for $r<R_{{\rm min},N}(T)$, i.e., $E_{\rm rep} =
A_{\rm rep} \theta[R_{{\rm min},N}(T) - r]$ with $A_{\rm rep} \gg
A_N$. The effective binding energy is then given by $E_{\rm eff} =
E_{{\rm Au}N} + E_{\rm rep}$. In the case of the investigated  
icosahedral structure the values are $R_{{\rm min},55}(T_{\rm room})
\approx 0.4$ nm and $R_{{\rm min},147}(T_{\rm room}) \approx 0.6$
nm. Note that larger clusters are structurally more stable than
smaller clusters and consequently the fusion process at $k_{\rm B} T =
0.0001$ eV ($T = 1.16$ K) leaves the initial structure of the two
Au$_{147}$ clusters more intact [Fig.\ \ref{clusterpot}(b)] than in
the case of two Au$_{55}$ clusters [Fig.\ \ref{clusterpot}(a)].

\subsection{Shape and adsorption energy of the adsorbed
  clusters}\label{sectionAdsorbed}

For the interpretation of the holographic analysis of the Au film and
the Au clusters it is useful to have information about their
shape. Moreover, from the simulations we obtain the effective
attractive potential of the substrate exerted onto clusters of
different size.

\subsubsection{Amorphous carbon substrate}

\begin{figure}
     \centerline{
       \includegraphics*[width=0.475\textwidth,clip=true]
       {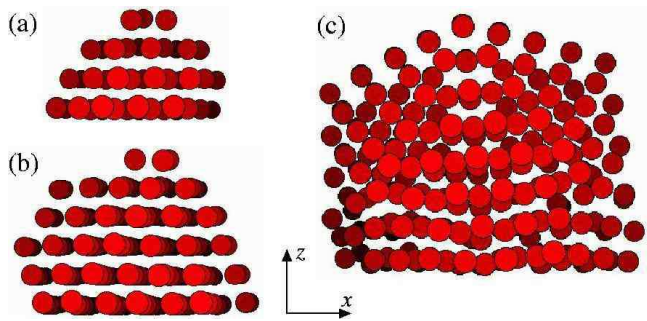}}
     \caption{\label{clustershape}\sl (Color online)
       Side view of the shape of adsorbed clusters of size (a) $N=55$,
       (b) $N=147$, and (c) $N=309$. The initially icosahedral
       clusters are placed on a planar substrate in the $x$-$y$ plane
       modeled with a generalized Lennard-Jones 
       potential Eq.\ \protect{\ref{LJfitmean}} with a depth of the
       minimum of $V_{{\rm LJ},0} = -0.3$ eV and simulated at room
       temperature ($k_{\rm B}T = 0.0256$ eV). The clusters are
       reconstructed in hcp (a), fcc (b), and distorted icosahedral
       (c) structures as a consequence of the boundary condition
       imposed by the substrate. (The images were taken after
       quenching to $T=1.16$ K in order to eliminate noise.)}     
   \end{figure}

Since the exact shape of the substrate surface is not known (c.f.\
discussion in Sec.\ \ref{sectionsurface}) we simulate the shape of
clusters for flat surfaces with different relevant effective mean
substrate-adsorbate potentials derived in Sec.\
\ref{SubstrateDFT}. Figure \ref{clustershape} shows a side view the
shape of adsorbed clusters (substrate at the bottom) in the presence
of a flat substrate with an attractive potential as given by Eq.\
\ref{LJfitmean} with a depth of $V_{{\rm LJ},0} = 0.3$ eV. The
clusters of size (a) $N=55$, (b) $N=147$, and (c) $N=309$ were
prepared in icosahedral configurations close to the surface and
subsequently freely evolved in a simulation run at room temperature
($k_{\rm B}T = 0.0256$ eV) for 10$^7$ to 10$^8$ Monte-Carlo steps per
atom.  

As a result of the boundary condition imposed by the substrate the
clusters reconstructed in the simulation process to hcp (a), fcc (b),
and distorted icosahedral (c) structures. As expected, the larger
$N=309$ cluster is more stable than the smaller ones and does not
undergo a structural transition within reasonable run times. The
energy gain due to the Au-substrate interaction is roughly the number
of atoms at the surface times the potential depth $V_{{\rm LJ},0}$. To
be specific, we find $\Delta E_{{\rm subs},55} = -6.31$ eV, $\Delta
E_{{\rm subs},147} = -10.85$ eV, and $\Delta E_{{\rm subs},309} =
-12.02$ eV for the configurations shown in Fig.\ \ref{clustershape} as
compared to the icosahedral structures.

When briefly tempering the clusters with $N=55$ and $N=147$ at $k_{\rm
B}T\sim 0.7$ eV and subsequent quenching we find that both hcp and
fcc structures are realized. The energy differences of the hcp
configuration with respect to the bulk-ground-state fcc structure is
smaller than energy fluctuations due to boundary effects. This
observation is consistent with the frequently observed\cite{CW97}
stacking faults in fcc and hcp bulk crystals.

For completeness we have performed runs simulating a graphite
substrate using Eq.\ (\ref{LJfitgraphite}) with a potential depth of
$V_{{\rm LJ},0} = 0.09$ eV. We find that Au$_{55}$ is amorphous at
room temperature while  Au$_{147}$ retains its initial icosahedral
structure (not shown). The melting point\cite{Wern05a} of Au$_{55}$ in
the absence of the substrate is close to room
temperature,\cite{EAT91,CLL99,CB01,KIS04,WernUn} which accounts for
its larger sensitivity to boundary effects.

\subsubsection{Gold film}

When placing a number of randomly distributed Au atoms on the
homogeneous substrate as described by Eq.\ (\ref{LJfitmean}) a
hexagonally coordinated double layer is found to be the metastable
structure attained (see also Sec.\ \ref{sectionDrag}) for a large 
parameter range of $0.1 < V_{{\rm LJ},0} / {\rm eV} < 0.6$. The
surface of such an Au double layer corresponds to an Au(111) surface,
albeit with 3.5\% contracted nearest-neighbor bond-length due to
surface effects.\cite{Gupt81} In order to model the shape of the Au
clusters that are found on the Au film we placed an initially
icosahedral cluster $N=309$ on the substrate surrounded by such an Au 
double layer formed of 291 atoms. The diameter of such a cluster of
$d\approx2.1$ nm corresponds to that investigated in Sec.\
\ref{sectionThick}.  

\begin{figure}
     \centerline{
       \includegraphics*[width=0.475\textwidth,clip=true]
       {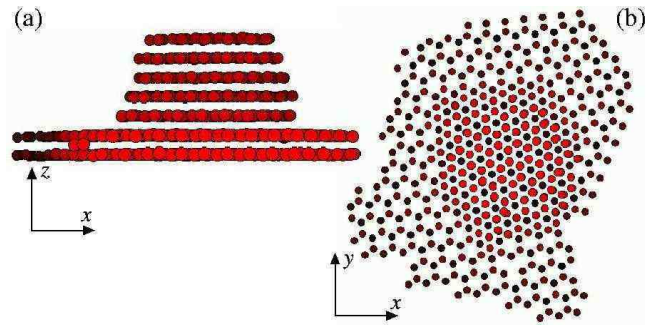}}
     \caption{\label{ClusterShape309}\sl (Color online)
       Side (a) and top (b) view of an Au$_{309}$ cluster surrounded
       by an Au double layer bound to a substrate as modeled by Eq.\ 
       (\protect\ref{LJfitmean}) with $V_{{\rm LJ},0} = 0.3$ eV at
       room temperature ($k_{\rm B}T = 0.0256$ eV). The Au$_{309}$ was
       initially icosahedral but reconstructed to a fcc lattice under
       the influence of the Au(111) double layer. (The images were
       taken after quenching to $T=1.16$ K in order to eliminate
       noise.)} 
   \end{figure}

Figure \ref{ClusterShape309} shows the resulting shape of the cluster
from the side in panel (a) and from the top in panel (b). Since the
Au-double-layer-cluster interaction is much stronger than the
substrate-cluster interaction [c.f.\ Fig.\ \ref{clustershape}(c)], the
cluster completely reconstructs to an fcc lattice to match the
structure of the film.

\subsection{Island formation}\label{sectionMCislands}

In order to model the formation of the islands on the a-C substrate
we simulate the dynamics of clusters rather than individual
atoms, which allows us to model sufficiently large systems. Moreover,
we are able to reproduce experimental results\cite{PKW+01,EHD02,CPP03}
for the diffusion and agglomeration of size selected clusters
published previously.

In Sec.\ \ref{sectionThick} we have shown from the experimental
holographic data that the Au film in the islands is less than 7 nm. In
Sec.\ \ref{sectionAdsorbed} the MC simulations have shown that small
adsorbed clusters of Au$_N$ for $N=55$ and  $N=147$ do not retain their
structure at room temperature and can easily fuse. Smaller clusters
are more likely to be trapped by pinning centers on the corrugated a-C
film as discussed in Secs.\ \ref{SubstrateHolography} and
\ref{SubstrateDFT}, while larger clusters are more stable and retain
their structure instead of fusing with their environment as indicated
in Secs.\ \ref{sectionClusterPot} and \ref{sectionAdsorbed}. These
findings suggest that the Au films of the islands are formed by
diffusing and coalescing clusters on the substrate in the size range
around $50 < N < 300$ leading to a film thickness of 4 to 8
monolayers. 

For simplicity we show here simulations of a system with an effective
cluster cohesive potential of the form given by Eq.\
(\ref{ClustercohesiveE}) with parameters as obtained for $N =
147$. Correspondingly, the cluster-substrate attractive potential is
given by Eq.\ (\ref{LJfitmean}), where we use a value of $V_{\rm LJ,0}
= 11$ eV as obtained from the quantitative analysis in Sec.\
\ref{sectionAdsorbed}. Results obtained for parameters with $N = 55$
are qualitatively equivalent (not shown).

\begin{figure}
     \centerline{
       \includegraphics*[width=0.475\textwidth,clip=true]
       {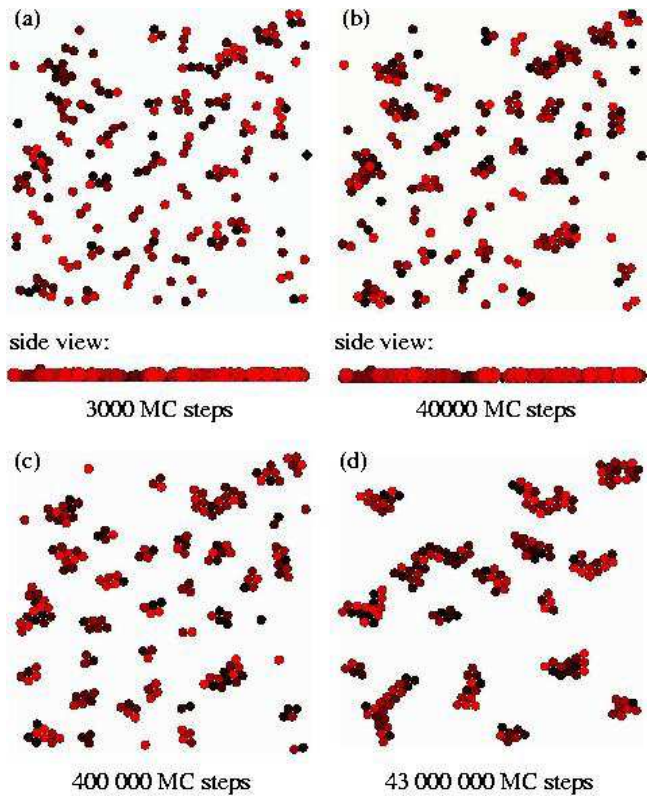}}
     \caption{\label{Islands147}\sl (Color online)
       Simulation of 200 Au$_{147}$ clusters interacting as described
       by Eq.\ (\protect\ref{ClustercohesiveE}) on a substrate
       modeled by Eq.\ (\protect\ref{LJfitmean}) with $V_{\rm LJ,0} =
       11$ eV after (a) 3000, (b) $4\times 10^{4}$, (c) $4\times
       10^{5}$, and (d) $4.3\times 10^{7}$ MC steps per cluster. The
       radius of the dots representing the clusters has been chosen as
       $R_{{\rm min},147}(T_{\rm room}) = 0.6$ nm, i.e., touching
       clusters are fused as shown in Fig.\
       \protect\ref{clusterpot}(b). The side views in panels (a) and
       (b) show that all clusters are located in a plane touching
       the substrate.\protect\cite{colors}}     
   \end{figure}

Figure \ref{Islands147} shows a series of snapshots\cite{colors} from
a simulation run of 200 initially randomly placed effective Au$_{147}$
clusters. Panel (a) shows that after 3000 MC steps per cluster small
islands and chains of clusters have formed very similar to those
observed on short time scales for monodisperse 4.8 nm diameter Au
clusters on an a-C substrate,\cite{EHD02} Ag$_{5000}$ clusters on
graphite,\cite{CPP03} and 8.5 nm diameter Co clusters on microgrid
substrates.\cite{PKW+01} It is important to note that the time scales
for the diffusion and fusion processes depend on the temperature, the
substrate properties, and the cluster sizes so that these metastable
structures can be observed only in a system specific time window.

At room temperature the Au$_{147}$ clusters can be considered as to be
fused almost instantaneously compared with the diffusion time scales
as shown in Fig.\ \ref{clusterpot}(b). Correspondingly the radius of
the dots representing the Au$_{147}$ clusters in Fig.\
\ref{Islands147} has been chosen as $R_{{\rm min},147}(T_{\rm room})
= 0.6$ nm. Larger clusters fuse on much longer time scales or at
higher temperatures.\cite{PKW+01,EHD02,CPP03,LJB97,ABPG04} From the MC
simulations a diffusion coefficient cannot be determined
quantitatively because of the lack of a time constant in the MC
procedure.

Panels (b), (c) and (d) of Fig.\ \ref{Islands147} shows the system
after $4\times 10^{4}$, $4\times 10^{5}$, and $4.3\times 10^{7}$ MC
steps per cluster, respectively. Clusters and islands of clusters
coalesce once they get close enough gaining cohesive energy. Structures
with short border lines are favored over extended linear ones for the
same reason. The (expected) tendency to form larger islands by
diffusion and fusion is clearly visible. 
The present simulation of effective clusters underestimates the
migration speed as well as the reorganization\cite{DMH81} of the
islands to more circular structures because the diffusion of the
individual atoms on the surface of the clusters is not accounted
for. For larger clusters the latter is the dominant cluster migration
process\cite{Grub67,WG75} referred to as surface diffusion. Including
this effects requires the investigation of much smaller systems such
as shown in Sec.\ \ref{sectionDrag}.   

In conclusion the cluster diffusion and fusion as modeled by the
effective clusters and cluster interactions correctly accounts for the 
experimentally observed\cite{EHD02,CPP03,PKW+01} pattern formation
[Fig.\ \ref{Islands147}(a)]. The fused clusters [Fig.\
\ref{clusterpot}(b)] form small islands [Fig.\ \ref{Islands147}(d)]
with a thickness of roughly the cluster diameter, which corresponds in
the case of Au$_{147}$ to 5 monolayers [Fig.\
\ref{clustershape}(b)]. The expected continued migration by surface 
diffusion\cite{Grub67,WG75} and fusion of the small islands in Fig.\
\ref{Islands147}(d) is consistent with the experimentally observed
island formation shown in Fig.\ \ref{expfig1}(b).

\subsection{Cluster drag}\label{sectionDrag}

All initially randomly distributed larger clusters are incorporated
into the Au films in the islands as shown in Figs.\ \ref{expfig1}(b)
and \ref{expfig2}(a).  It is not possible to simulate this effect
directly because of the size limitations of the method. It is possible
though to place randomly distributed Au atoms on a substrate modeled by
the generalized Lennard Jones potential Eq.\ (\ref{LJfitmean}), where
$V_{{\rm LJ},0} = 0.3$ eV, together with a $N=147$ and a $N=55$ Au
cluster at room temperature ($k_{\rm B}T = 0.0256$ eV). The thus
modeled system is significantly smaller than those observed in the
experiments but the dynamical behavior is likely to be similar, albeit
with rescaled diffusion constants and time scales.

\begin{figure}
     \centerline{
       \includegraphics*[width=0.475\textwidth,clip=true]
       {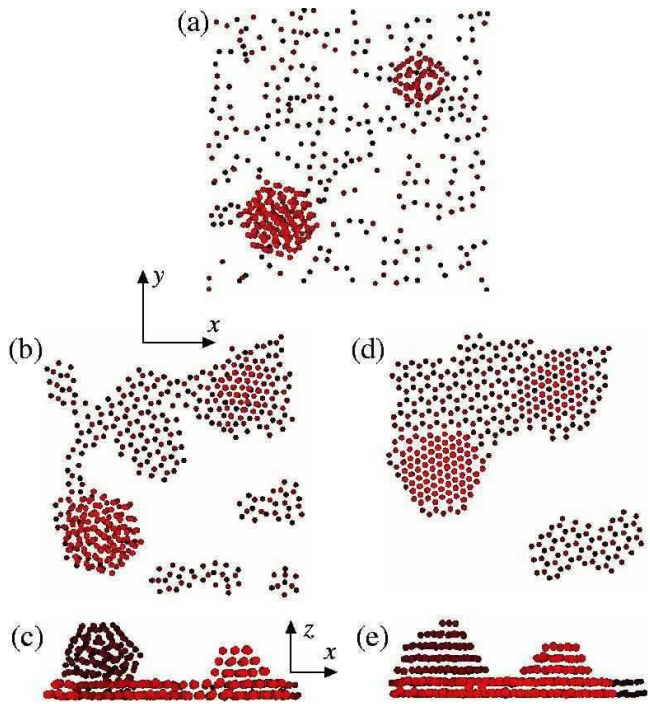}}
     \caption{\label{ClusterDrag}\sl (Color online)
       Time evolution of a system of 298 Au atoms placed randomly on
       the substrate [Eq.\ (\ref{LJfitmean})], one Au$_{\rm 55}$
       and one Au$_{\rm 147}$ cluster. Panel (a): 93 MC steps; panels (b) 
       and (c): $6\times10^{4}$ MC steps; panels (d) and (e):
       $1.18\times10^{6}$ MC steps per atom. Panels (c) shows the
       formation of the double layer film, panel (e) the restructuring
       of the initially icosahedral clusters. (The images were taken
       after quenching to $T=1.16$ K in order to eleminate noise.)}  
   \end{figure}

Snapshots taken at three different run times are shown in Fig.\
\ref{ClusterDrag}. Panel (a) shows the initially random distribution
of 298 atoms as well as the Au$_{\rm 55}$ and Au$_{\rm 147}$ clusters
after 93 MC steps per atom. Panel (b) shows the system after
$6\times10^{4}$ MC steps per atom. The percolating regions
coalesce. Panel (c) shows a side view of the same state of the system
as in (b) and illustrates that the atoms have formed a Au double
layer. This double layer is stable at room temperature in a substrate
parameter range of $0.1 < V_{{\rm LJ},0}/{\rm eV} < 0.6$.

Panel (d) of Fig.\ \ref{ClusterDrag} shows the system after
$1.18\times10^{6}$ MC steps per atom. The coalescing Au double layer
has contracted thus minimizing it cohesive energy. In the cohesion
process the less mobile Au$_{\rm 55}$ and Au$_{\rm 147}$ clusters have
been pulled closer together. They are located closer to the border of
the Au film consistent with the experiments indicating a slightly
larger probability to find clusters near the boundaries of the islands
than in their center. The separated island in the lower right hand
corner of Fig.\ \ref{ClusterDrag} is less mobile and fuses with the
larger island only after comparably long run times (not shown).

Panel (e) of Fig.\ \ref{ClusterDrag} shows a side view of the system
after $1.18\times10^{6}$ MC steps per atom and illustrates the fcc
restructuring of the initially icosahedral clusters in the presence of
the Au double layer film.


\section{Conclusions}\label{sectionConclusions}

We presented experimental results showing the formation of Au islands
on amorphous carbon substrates after laser deposition of
non-size-selected clusters and subsequent aging of the samples in the 
absence of inert condition for three to four months. We characterized
both the substrate and the Au films with electron transmission
holography. A model potential for the substrate-Au interaction was
derived with the help of density functional calculations. Subsequently
a number of Monte-Carlo simulations were carried out describing the
island formation process and the shape of the adsorbed clusters.

The following qualitative and quantitative results have been incurred
from the investigations.

\begin{itemize}
\item The observed islands have a typical size of  $A \sim 25 \times
  70$ nm$^2$.
\item The islands consist of an Au film of a few monolayers [probably
  4 to 8 but definitely less than 27 Au(111) layers] thickness and
  adsorbed larger clusters. 
\item The islands are metastable and coalesce at elevated temperatures
  of $T \sim 400$ K.
\item The islands are formed by diffusion and coalescence of clusters
  of size $50 < N < 300$.
\item Larger clusters, which normally are less mobile, are dragged
  along by the percolating Au film formed by the smaller clusters.
\item Smaller clusters are pinned at deeper substrate intrusions as
  indicated by the broadening of the phase distribution in the
  holographic images of the ablated samples.
\item A number of observations suggest that the amorphous carbon
  substrate surface is much less corrugated than anticipated. Instead,
  the substrate appears to exhibit significant internal spatial density
  fluctuations.
\item The Au-substrate interaction can be modeled by a generalized
  Lennard-Jones potential for graphite [Eq.\ (\ref{LJfitgraphite})]
  and for amorphous carbon [Eq.\ (\ref{LJfitmean})]. 
\item An effective Au cluster-cluster interaction has been determined
  [Eq.\ (\ref{ClustercohesiveE})].  
\end{itemize}

The extensive investigation presented in this work contribute to a
better understanding of some processes taking place in the so far
little investigated field of the dynamic behavior of cluster arrays
deposited on amorphous surfaces.


\begin{acknowledgments}

We thank S.-S.\ Jester and M.\ M.\ Kappes for the sample preparation,
outlining the preparation process, and discussions. RW thanks
P.\ Schmitteckert, M.\ Vojta, and P.\ W\"olfle for instructive
discussions. The work was supported by the Center for Functional
Nano\-struc\-tures of the Deutsche Forschungsgemeinschaft within
project area D1.

\end{acknowledgments}


\begin{thebibliography}{10}

\bibitem{WG75}
P. Wynblatt and N.~A. Gjostein, Progr. Sol. State Chem. {\bf 9},  21  (1975).

\bibitem{Will87}
P. Williams, Appl. Phys. Lett. {\bf 50},  1760  (1987).

\bibitem{MRC96}
K. Morgenstern, G. Rosenfeld, and G. Comsa, Phys. Rev. Lett. {\bf 76},  2113
  (1996).

\bibitem{PKW+01}
D.~L. Penga, T.~J. Konno, K. Wakoh, T. Hihara, and K. Sumiyama, Appl. Phys.
  Lett. {\bf 78},  1535  (2001).

\bibitem{EHD02}
D.~A. Eastham, B. Hamilton, and P.~M. Denby, Nanotechnology {\bf 13},  51
  (2002).

\bibitem{CPP03}
M. Couillard, S. Pratontep, and R.~E. Palmera, Appl. Phys. Lett. {\bf 82},
  2595  (2003).

\bibitem{FSCN04}
M. Forshaw, R. Stadler, D. Crawley, and K. Nikolic, Nanotechnology {\bf 15},
  220  (2004).

\bibitem{ADKL02}
G. Allan, C. Delerue, C. Krzeminski, and M. Lannooi, Nanostruct. Mat.  161
  (2002).

\bibitem{PLK+03}
A. Pestryakov, V. Lunin, A. Kharlanov, N. Bogdanchikova, and I. Tuzovskaya,
  Euro. Phys. J. D {\bf 24},  307  (2003).

\bibitem{IAO+03}
S. Ichikawa, T. Akita, M. Okumura, M. Kohyama, and K. Tanaka, JEOL News {\bf
  38},  6  (2003).

\bibitem{Ostw00}
W. Ostwald, Z. Phys. Chem. (Leipzig) {\bf 34},  495  (1900).

\bibitem{Wagn61}
C. Wagner, Z. Elektrochem. {\bf 65},  581  (1961).

\bibitem{Utla80}
M. Utlaut, Phys. Rev. B {\bf 22},  4650  (1980).

\bibitem{SPWB86}
D.~J. Smith, A.~K. Petford-Long, L.~R. Wallenberg, and J.-O. Bovin, Science
  {\bf 233},  872  (1986).

\bibitem{ESZ+00}
M.~Y. Efremov, F. Schiettekatte, M. Zhang, E.~A. Olson, A.~T. Kwan, R.~S.
  Berry, and L.~H. Allen, Phys. Rev. Lett. {\bf 85},  3560  (2000).

\bibitem{WDM+00}
D.~J. Wales, J.~P.~K. Doye, M.~A. Miller, P.~N. Mortenson, and T.~R. Walsh,
  Adv. Chem. Phys. {\bf 115},  1  (2000).

\bibitem{HSH+03}
J. Hagen, L. Socaciu, U. Heiz, T. Bernhardt, and L. W\"oste, Euro. Phys. J. D
  {\bf 24},  327  (2003).

\bibitem{Gros01}
D. Gross, {\em Microcanonical thermodynamics: Phase transitions in Small
  systems}, {\em volume 66 of Lecture Notes in Physics} (World Scientific,
  Singapore, 2001).

\bibitem{FGH+04}
K. Fauth, S. Gold, M. He{\ss}ler, N. Schneider, and G. Sch\"utz, Chem. Phys.
  Lett. {\bf 392},  498  (2004).

\bibitem{VMG03}
W. Vervisch, C. Mottet, and J. Goniakowski, Euro. Phys. J. D {\bf 24},  311
  (2003).

\bibitem{Reim89}
L. Reimer, {\em Transmission electron microscopy} (Springer, Berlin, 1989).

\bibitem{LL02}
M. Lehmann and H. Lichte, Microsc. Microanal. {\bf 8},  447  (2002).

\bibitem{Robe86}
J. Robertson, Adv. Phys. {\bf 35},  317  (1986).

\bibitem{RFG04}
B. Reznik, M. Fotouhi, and D. Gerthsen, Carbon {\bf 42},  1305  (2004).

\bibitem{TL04}
J.~T. Titantah and D. Lamoen, Phys. Rev. B {\bf 70},  033101  (2004).

\bibitem{WGGK02}
P. Weis, S. Gilb, P. Gerhardt, and M.~M. Kappes, Int. J. Mass Spec. {\bf 216},
  59  (2002).

\bibitem{WWVK04}
P. Weis, O. Welz, E. Vollmer, and M.~M. Kappes, J. Chem. Phys. {\bf 120},  677
  (2004).

\bibitem{MdH90}
P. Milani and W.~A. deHeer, Rev. Sci. Instrum. {\bf 61},  1835  (1990).

\bibitem{DDPS81}
T.~G. Dietz, M.~A. Duncan, D.~E. Powers, and R.~E. Smalley, J. Chem. Phys. {\bf
  74},  6511  (1981).

\bibitem{HVTS97}
U. Heiz, F. Vanolli, L. Trento, and W.-D. Schneider, Rev. Sci. Instrum. {\bf
  68},  1986  (1997), and references therein.

\bibitem{RKR+96}
A. Rosenauer, S. Kaiser, T. Reisinger, J. Zweck, W. Gebhardt, and D. Gerthsen,
  Optik {\bf 102},  63  (1996).

\bibitem{Phasesign}
Since the wavelength of the electronic wave-function is longer in the sample
  medium than in vacuum \protect\cite{Reim89} the phase shift measured is
  actually negative. For a more intuitive representation, where larger values
  correspond to larger sample thickness, we have inverted the $\phi$ axes.

\bibitem{Buhl59}
R. Buhl, Z. Phys. {\bf 155},  395  (1959).

\bibitem{Kell61}
M. Keller, Z. Phys. {\bf 164},  274  (1961).

\bibitem{GC99}
M. Gajdardziska-Josifovska and A.~H. Carim,  in {\em Introduction to Electron
  Holography}, edited by E. Voelkl, L.~F. Allard, and D.~C. Joy (Plenum Press,
  New York and London, 1999), pp.\ 241--268.

\bibitem{HL98}
A. Harscher and H. Lichte,  in {\em Proceedings of the 14th International
  Congress on Electron Microscopy}, edited by H. Calderon-Benavides and M.~J.
  Jose-Yacaman (IOP Publishing Ltd., London, 1998), Vol.~1, p.\ 553.

\bibitem{SO85}
A. S\'anchez and M. Ochando, J. Phys. C: Solid St. Phys. {\bf 18},  33  (1985).

\bibitem{RB98}
S.~M. Ritzau and R.~A. Baragiola, Phys. Rev. B {\bf 58},  2529  (1998).

\bibitem{AWWB03}
F. Allegrini, R.~F. Wimmer-Schweingruber, P. Wurz, and P. Bochsler, Nucl.
  Instr. and Meth. B {\bf 211},  487  (2003).

\bibitem{WP91}
Y. Wang and J.~P. Perdew, Phys. Rev. B {\bf 44},  13298  (1991).

\bibitem{Bloe94}
P.~E. Bl\"ochl, Phys. Rev. B {\bf 50},  17953  (1994).

\bibitem{KJ99}
G. Kresse and D. Joubert, Phys. Rev. B {\bf 59},  1758  (1999).

\bibitem{DFTquote}
A planwave cutoff of E$_{\rm cut}$=275eV was used, with additional calculations
  at E$_{\rm cut}$=400eV to check convergence. K-point sampling on a $8\times
  8$ ($2\times 2$) grid in the 2 dimensional Brillouin zone was used for the
  graphite (a-C) calculations, respectively. Total energies were converged to
  $10^{-5}$ eV to ensure accurate forces.

\bibitem{AT89}
M.~P. Allen and D.~J. Tildesley, {\em Computer Simulation of Liquids}, {\em
  Oxford Science Publications} (Clarendon Press, Oxford, 1989).

\bibitem{WDS+01}
J. Wanga, F. Dinga, W. Shena, T. Lia, G. Wanga, and J. Zhaod, Solid State
  Commun. {\bf 119},  13  (2001).

\bibitem{Wern05a}
R. Werner, Eur.\ Phys. J.\ B  in print  (2005).

\bibitem{Gupt81}
R.~P. Gupta, Phys. Rev. B {\bf 23},  6265  (1981).

\bibitem{CR93}
F. Cleri and V. Rosato, Phys. Rev. B {\bf 48},  22  (1993).

\bibitem{HBB+97}
H. H\"ovel, T. Becker, A. Bettac, B. Reihl, M. Tschudy, and E.~J. Williams, J.
  Appl. Phys. {\bf 81},  154  (1997).

\bibitem{Hoev01}
H. H\"ovel, Appl. Phys. A {\bf 72},  295  (2001).

\bibitem{DD98}
J. Dong and D.~A. Drabold, Phys. Rev. B {\bf 57},  15591  (1998).

\bibitem{Urba98}
J. Urban, Cryst. Res. Technol. {\bf 33},  1009  (1998).

\bibitem{GWF+02}
S. Gilb, P. Weis, F. Furche, R. Ahlrichs, and M.~M. Kappes, J. Chem. Phys. {\bf
  116},  4094  (2002).

\bibitem{McLe69}
R.~B. McLellan, Scr. Metall. {\bf 3},  389  (1969).

\bibitem{HCR+87}
V.~M. Hallmark, S. Chiang, J.~F. Rabolt, J.~D. Swalen, and R.~J. Wilson, Phys.
  Rev. Lett. {\bf 59},  2879  (1987).

\bibitem{HCZC95}
L. Huang, J. Chevrier, P. Zeppenfeld, and G. Cosma, Appl. Phys. Lett. {\bf 66},
   935  (1995).

\bibitem{HKS+81}
M.~A. van Hove, R.~J. Koestner, P.~C. Stair, J.~P. Biberian, L.~L. Kesmodel, I.
  Bartos, and G.~A. Somorjai, Surf. Sci. {\bf 103},  189  (1981).

\bibitem{GSC88}
E. Ganz, K. Sattler, and J. Clarke, Phys. Rev. Lett. {\bf 60},  1856  (1988).

\bibitem{DMH81}
M. Drechsler, J.~J. M\'etois, and J.~C. Heyraud, Surf. Sci. {\bf 108},  549
  (1981).

\bibitem{LJB97}
L.~J. Lewis, P. Jensen, and J.-L. Barrat, Phys. Rev. B {\bf 56},  2248  (1997).

\bibitem{ABPG04}
S. Arcidiacono, N. Bieri, D. Poulikakos, and C. Grigoropoulos, Int. J. Multiph.
  Flow {\bf 30},  979  (2004).

\bibitem{HCSR97}
O.~D. H\"aberlen, S.-C. Chung, M. Stener, and N. R\"osch, J. Chem. Phys. {\bf
  106},  5189  (1997).

\bibitem{GMB+98}
I.~L. Garz\'on, K. Michaelian, M.~R. Beltr\'an, A. Posada-Amarillas, P.
  Ordej\'on, E. Artacho, D. S\'anchez-Portal, and J.~M. Soler, Phys. Rev. Lett.
  {\bf 81},  1600  (1998).

\bibitem{DW98b}
J.~P.~K. Doye and D.~J. Wales, Phys. Rev. Lett. {\bf 80},  1357  (1998).

\bibitem{CW97}
N. Chetty and M. Weinert, Phys. Rev. B {\bf 56},  10844  (1997).

\bibitem{EAT91}
F. Ercolessi, W. Andreoni, and E. Tosatti, Phys. Rev. Lett. {\bf 66},  911
  (1991).

\bibitem{CB01}
Y.~G. Chushak and L.~S. Bartell, J. Phys. C {\bf 105},  11605  (2001).

\bibitem{KIS04}
K. Koga, T. Ikeshoji, and K.~I. Sugawara, Phys. Rev. Lett. {\bf 92},  115507
  (2004).

\bibitem{CLL99}
C.~L. Cleveland, W.~D. Luedtke, and U. Landman, Phys. Rev. B {\bf 60},  5065
  (1999).

\bibitem{WernUn}
R.\ Werner, unpublished.

\bibitem{colors}
The color coding of Figs.\ 10 through 13 is such that the objects closest to
  the viewer are lighter and the most distant objects are dark. The color
  gradient is normalized to the maximum difference in distance of objects. As a
  result the different clusters shown in Fig.\ 12 appear in different shades
  even though they all are positioned in the same plane with very little
  difference in distance from the substrate, which becomes apparent in the side
  views in panel (a) and (b).

\bibitem{Grub67}
E.~E. Gruber, J. Appl. Phys. {\bf 38},  243  (1967).

\end{thebibliography}

\end{document}